# Astro2020 Science White Paper

# High-Energy Astrophysics in the 2020s and Beyond

**Thematic Areas:** ☐ Planetary Systems   ☒ Star and Planet Formation
☒ Formation and Evolution of Compact Objects   ☒ Cosmology and Fundamental Physics
☒ Stars and Stellar Evolution   ☒ Resolved Stellar Populations and their Environments
☒ Galaxy Evolution   ☒ Multi-Messenger Astronomy and Astrophysics

**Principal Author:**
Name: Christopher Reynolds
Institution: University of Cambridge, United Kingdom
Email: csr12@ast.cam.ac.uk
Phone: +44-1223-766668

**Co-authors:**
*Executive Committee of the AAS High-Energy Astrophysics Division (HEAD):*
Rob Petre (NASA-GSFC)                     Kristin Madsen (Caltech)
Michael Corcoran (USRA)                   Gabriela Gonzales (LSU)
Keith Arnaud (U.Maryland)                 Laura Brenneman (Harvard-Smithsonian CfA)
Niel Brandt (Penn State Univ.)            Laura Lopez (Ohio State Univ.)
Neil Cornish (Montana State)

**Abstract:**
With each passing decade, we gain new appreciation for the dynamic, connected, and often violent nature of the Universe. This reality necessarily places the study of high-energy processes at the very heart of modern astrophysics. This White Paper illustrates the central role of high-energy astrophysics to some of the most pressing astrophysical problems of our time, the formation/evolution of galaxies, the origin of the heavy elements, star and planet formation, the emergence of life on exoplanets, and the search for new physics. We also highlight the new connections that are growing between astrophysicists and plasma physicists. We end with a discussion of the challenges that must be addressed to realize the potential of these connections, including the need for integrated planning across physics and astronomy programs in multiple agencies, and the need to foster the creativity and career aspirations of individual scientists in this era of large projects.

## The Ubiquity of High-Energy Processes

The history of astronomy is defined by the increasing realization that the Universe is dynamic, violent, and often extreme. The study of the hot and energetic Universe, i.e., high energy astrophysics (HEA), now finds itself at the very core of modern astrophysics. Further, the new frontiers of multi-messenger astronomy (including recent dramatic advances in the gravitational wave and neutrino realms) and time-domain astronomy naturally fall under the HEA umbrella; the astrophysical objects driving these phenomena are the very same compact objects that have been the focus of many high-energy astrophysicists since the earliest days of the field.

In this White Paper, we highlight the connectedness of high-energy astrophysics to essentially every other key question in modern astrophysics. Our presentation is motivated by the Special HEAD Meeting on "High-Energy Astrophysics in the 2020s and Beyond" held in Rosemont, IL, in March-2018 at which the HEA community came together to discuss future directions and capabilities[1]. Here, we do not enter into discussions of specific future missions or projects, leaving that to other Astro2020 submissions. The purpose of this current White Paper is to contextualize the numerous HEA community Astro2020 submissions into a wider framework.

## A Universe of Black Holes

Black holes used to be considered cosmic oddities but are now recognized to be one of *the* critical players in many realms of astrophysics. Here we highlight just a few areas in which black holes find themselves at the center of contemporary astrophysics, and the role that high-energy observations have to play in in the future of these fields.

Supermassive black holes (SMBHs) are now widely accepted to dominate feedback and regulation processes in the evolution of massive galaxies (Ruszkowski, 2019), but there are now indications that their influence may extend down to lower-mass and even dwarf systems (Gallo, 2019). This forces us to confront a difficult multi-scale problem, with the (still far from understood) details of black hole accretion physics on sub-parsec scales having a defining role in shaping galactic-scale structure. For example, we do not yet have a robust understanding of the partitioning of the luminosity of active galactic nuclei (AGN) into radiative luminosity, sub-relativistic winds (Laha, 2019; Tombesi, 2019), and highly-relativistic jets (Perlman, 2019). As impressive as theoretical progress has been, even today's state-of-the-art cosmological structure formation simulations need to make the crudest of assumptions about the inner workings of the AGN. So AGN feedback will remain an observationally driven field for the foreseeable future, with X-ray studies of the SMBH and AGN central engine (Zoghbi, 2019; Kamraj, 2019; Tombesi, 2019; Laha, 2019; Garcia, 2019; Lopez-Rodriguez, 2019; Chartas, 2019) and radio/gamma-ray studies of jets and cosmic ray population (Perlman, 2019; Santander, 2019; Meyer, 2019) at the forefront. Accreting Galactic stellar mass black holes also present uniquely accessible laboratories for accretion physics (Miller, 2019; Maccarone, 2019).

Despite their importance to galaxy evolution, we still do not know the correct paradigm for the formation of SMBHs (Natarajan, 2019; Greene, 2019); were the first high-redshift seeds of SMBHs created by the core collapse of Population III stars, the monolithic collapse of massive gas clouds in the cores of protogalaxies, the dynamical collapse of a compact star cluster, or some yet to be conjectured pathway? The answer to this question has implications for the reionization of the Universe (Cooray, 2019; Basu-Zych, 2019) and the subsequent effects on early galaxy evolution. Next generation of X-ray observatories will allow SMBH spin measurements to be performed (Zoghbi, 2019; Garcia, 2019) out to high-redshift, providing a new window into early growth modes of these objects. Ultimately, though, progress requires the characterization of AGN at their formation redshifts, $z>10$, and deep X-ray surveys provide a clean probe of such objects (Fan, 2019). The launch of JWST and WFIRST in the coming decade will rapidly mature our knowledge of galaxies at these redshifts, laying the groundwork for future ultra-deep X-ray surveys enabled by breakthroughs in X-ray optics (Pacucci, 2019). Combining these future IR and X-ray capabilities, together with GW measurements of the SMBH merger history, will allow us to solve this 50-year old mystery within the next 20 years (Haiman, 2019).

---

[1] Presentations of this meeting are available at https://aas.org/meetings/high_energy_special_2018.

Staying with the theme of black hole formation, LIGO's detection of GW170817 and the subsequent electromagnetic follow-up has dramatically and conclusively demonstrated the link between black hole formation via the merger of neutron stars and the production of the heaviest elements in the Universe (Timmes, 2019). This connection dovetails with the realization that conditions within core-collapse supernovae are probably unable to sustain the r-process necessary for the production of these elements (Roederer, 2019; Binns, 2019), and highlights the importance of intensive spectroscopic studies of future events. Of course, spectroscopic follow-ups are only possible once the gravitational wave (GW) event has been localized. For GW170817, the *Fermi* Gamma-Ray Burst Monitor played an essential role in the initial localization. Future studies will rely upon continued developments in ground-based GW detectors together with all-sky monitors (ASMs) that can localize the electromagnetic counterpart within minutes. X-ray ASMs have a particularly important role to play in future GW localization experiments, because the X-ray flash/afterglow produced by a NS-NS merger is far brighter than most persistent X-ray sources and thus easily distinguished.

## The Hot Side of Galaxy Evolution

Half of all of the baryons in the Universe today are in the hot phase (T>1 million K): the warm/hot intergalactic medium, the intracluster medium (ICM) of galaxy clusters, and the hot circumgalactic medium (CGM) around field galaxies (Weisz, 2019). Here we highlight the importance of the CGM (Cicone, 2019; Zaritsky, 2019; Lebouteiller, 2019) and the role of future soft X-ray observations. Simulations suggest that the thermodynamic/multiphase structure and metallicity of the CGM is the most powerful marker of the AGN and stellar feedback physics that controls galaxy growth (Hodges-Kluck, 2019). It has even been suggested that the true quenching of galactic-scale star formation is driven by the heating of the hydrostatic CGM atmosphere and not simply the removal of molecular gas from the galactic disk. In this sense, there may be a continuity between the feedback physics that is well studied in the cluster/ICM context (Ruszkowski, 2019) and that occurring within individual galaxies.

Yet our current knowledge of the hot CGM is poor. CGM atmospheres are low-surface brightness and radiate predominantly in the soft X-ray band where foreground emission can be bright. As a result, there are still order-unity uncertainties in CGM temperatures, and order-of-magnitude uncertainties in densities, metallicities, and cool/cold gas content. The future will see rapid progress due to the advent of non-dispersive, high-spectral resolution X-ray detectors that can easily decompose the (spectrally rich and redshifted) CGM emission from the (rest-frame or featureless) X-ray foregrounds/background, and high spectral resolution dispersive X-ray spectrometers that can use bright quasars as backlight sources to measure the redshift and density of CGM structures. New maps of the thermal and kinetic Sunyaev-Zeldovich (SZ) effect enabled by future CMB experiments will also provide a highly complementary view of the CGM (Mroczkowski, 2019). Studies of the dramatic feedback event in the Milky Way, i.e. the Fermi Bubbles, also promise to be rewarding (Fox, 2019).

As is the case in their more massive cousins, galaxy clusters, the microphysics of the hot CGM is strongly dependent upon these densities and temperatures – densities can become low enough for the hot plasma to become only weakly collisional in which case poorly understood kinetic physics becomes relevant and can dictate how the feedback processes operate (Kunz, 2019; Markevitch, 2019). The recent appreciation for the need to understand weakly collisional plasmas is driving a burgeoning connection between astrophysicists interested in galactic hot baryons and those studying hot plasmas with in-situ experiments in the solar wind (Chen, 2019).

## The Non-Thermal Universe

Relativistic particles (aka cosmic rays; CR) are found in essentially every astrophysical setting, and there is an increasing realization that they play an important role in the energy and/or ionization budget of many systems. Recent work has highlighted the role that CRs play in the physics of feedback in galaxies (Ruszkowski, 2019). CR pressure may be the driving force behind stellar-driven galactic-scale winds, with radio and gamma-ray observations being the most direct probe of such physics (Orlando, 2019). The mixing and subsequent diffusion/streaming of CRs into the thermal ICM is also increasingly being cited as an important ingredient for cluster-scale feedback.

A fundamental understanding of the non-thermal Universe must start with a study of the CRs that we can detect directly. Despite being discovered a century ago, identifying the acceleration sites of CRs and determining the acceleration physics remains challenging (Cristofari, 2019; Schroeder, 2019; Vandenbroucke, 2019) – the charged nature of the particles curves their trajectory as they propagate through the Galactic magnetic field thereby erasing directional clues as to their origin. Gamma-rays and X-rays have therefore been used as proxies for CR acceleration. The role of supernova remnants in the Galactic CR production has indeed been verified with the detection of X-ray synchrotron emission emanating from TeV electrons in SNR shocks, and from the neutral pion-decay spectrum resulting from proton-proton collisions (Ackermann, 2013). However, for other sources, the inverse-Compton scattering of relativistic electrons may account for the observed gamma-ray spectrum leaving the accounting of the Galactic CR uncertain. It is particular interesting that some CRs may be left from past AGN activity in our Galaxy. (HESS, 2016)

## Breathing Life into Star and Planet Formation

We now understand that accretion disks are magnetohydrodynamic (MHD) systems and hence require at least some level of ionization. Thus, the accretion processes that assemble stellar and planetary systems fundamentally proceed due to X-ray and/or cosmic ray. The X-rays naturally originate from the young star itself, but the cosmic rays likely arise from external sources and may be give star formation a unique environmental dependence. X-rays and cosmic rays may play an equally profound role in driving the complex chemistry within protoplanetary disks that forms complex molecules, grains, and eventually planets themselves (Cleeves, 2017). Magnetic dynamo processes and reconnection may also play an essential role in these disks (Ji, 2019). Once an exoplanet forms, X-rays from the star can play an important role in the formation of the planetary atmosphere and the possible evolution of life (Wolk, 2019; Drake, 2019). Future progress will be built on collaborations between solar system researchers, the exoplanet community, cosmic ray researchers, and experts on meteorite paleomagnetism (as a probe of these processes in our own solar system).

Of course, much of the story of star and planet formation involves cold, molecular and dusty materials. X-ray astronomy provides highly complementary probes of this material. Dust scattering halos around bright Galactic X-ray sources probe the line-of-sight distribution of Galactic dust, and X-ray absorption spectroscopy probes the crystalline structure of grains via resonant substructure imprinted on the elemental photoelectric absorption edges (Valencic, 2019; Corrales, 2019). The X-ray emission from comets via the charge exchange process provides a future opportunity to constrain the early solar system elemental abundances as a function of location, and seek evidence for outflow events at the snowline. (Snios, 2019)

## The Universe as a Laboratory for Extreme Physics

The Universe provides us with a fertile ground for understanding the laws of nature in extreme situations. GWs produced by the collision of black holes, now detectable with ground-based observatories, are already providing tests of General Relativity in the very strong field limit (Sathyaprakash, 2019)– this coming decade will provide more precise tests as the detectors' sensitivities improve (Kalogera, 2019; Shoemaker, 2019). The future advent of space-based GW detectors will give exquisitely detailed signals, permitting an unprecedented view of strong gravity at work (Berti, 2019; Berry, 2019; Cornish, 2019; Tinto, 2019). We stress the very real possibility of there being a profound discovery in this field such as GW echoes from black hole firewalls, strong evidence for non-locality, or "fuzzy" black holes (Cornish, 2019); any such finding would reshape the foundations of theoretical physics for at least the next century.

The most obvious manifestation of physics beyond the standard model is dark matter. The failure to detect WIMPs, either directly in the laboratory or indirectly via astrophysical observations, is starting to stress thermal production mechanisms and is leading to a resurgence of interest in both fermionic (sterile neutrino) and bosonic (axion-like particle) alternatives. In any of these cases, astrophysical observations remain the most effective tool for searching for dark matter, with the X-ray (Kashlinsky, 2019) and gamma-ray bands (Caputo, 2019; Viana, 2019) being pre-eminent.

The highest energy cosmic rays and cosmic neutrinos far exceed energies attainable in terrestrial laboratories (Sarazin, 2019). The recent coincidence of an ICECUBE PeV neutrino event with enhanced gamma-ray activity of the blazar TXS 0506+056 firmly established the AGN and its relativistic jet as the likely source of the neutrinos and, by extension, the very high energy CRs that likely form them. This reinvigorates attempts to understand the physics of high-energy particle acceleration in AGN jets (Perlman, 2019; Rani, 2019; Santander, 2019)  A cherished foundation of physics, Lorentz invariance, will be tested with increasing precision in the decade by detection of extremely high energy cosmic rays and TeV gamma-rays from blazars and other astrophysical sources (Mukherjee, 2019; Vieregg, 2019). This is another opportunity for a profound discovery – violation Lorentz invariance will destroy a pillar of modern physics and produce insights to the quantum nature of spacetime.

X-ray observations offer the clearest window for studying the extremes of magnetism and density that are unique to neutron stars (Wolff, 2019; Wadiasingh, 2019). Existing, planned, and future X-ray polarimetry and high-resolution, high-throughput timing instruments will measure the neutron star equation of state, providing insight into nuclear processes unavailable from a terrestrial laboratory, and potentially reveal such elusive quantum phenomena as vacuum polarization (Krawczynski, 2019).

## Realizing Future Potential : Challenges and Opportunity

The principal message of this White Paper is that high-energy processes are truly ubiquitous throughout the Universe – high-energy astrophysics is no longer a distinct, niche, discipline. We simply do not have a complete view of the Universe if we are not folding in data from X-ray, gamma-ray, GW, particle, and neutrino observatories.  With the recognition that much of the Universe is in the plasma state, there connections between astrophysics and both theoretical and experimental plasma physicists are also growing at an exponential rate.  For these reasons, HEA is and will continue to be supported from both the Astronomy and Physics programs at NSF, and has a footprint in the Heliophysics as well as the Astrophysics Divisions at NASA. To maximize the science return, especially as more financially demanding facilities are needed for continued progress, there is a need for integrated planning across agency astronomy programs and between physics and astronomy/astrophysics. Further, coordination with our major international partners (whose planning cycle is disjoint from ours, leading to program overlap and unnecessary competition), especially on large programs, would lead to implementation of a wider variety of programs or missions.

Within the broad range of investigations, some are best performed by individuals or small teams, others by modest sized groups, and still others only by large consortia.  As the national agencies create their strategic plans, it is essential that they recognize that such a mix is healthy for the field and work diligently to maintain a balance among large, medium, and small programs both for space and ground-based facilities, and for the research not needing facilities (i.e., theory). At the same time, as a high proportion of resources inevitably is directed toward the large programs, it is important that these be structured so that program goals are balanced against creativity and career needs of the individual scientist.